\documentclass[prb,twocolumn,showpacs]{revtex4}
\usepackage{amsmath}
\usepackage{dcolumn}
\usepackage{graphicx}
\usepackage{physics}

\begin{document}

\title[Short title for running header]{The driving mechanism of the d-wave orbital order in the iron-based superconductors}
\author{Da-Wei Yao and Tao Li}
\affiliation{Department of Physics, Renmin University of China, Beijing 100872, P.R.China}
\date{\today}

\begin{abstract}
We study the driving mechanism and the form of the orbital order in the electronic nematic phase of the iron-based superconductors(IBSs) within the random phase approximation of a 5-band model. We find the magnetic correlation energy of the system can be significantly improved when an orbital order of the d-wave form is spontaneously generated. On the other hand, the magnetic correlation energy increases as one introduce either an on-site or an extended s-wave orbital order. More specifically, we find that the on-site orbital order is disfavored by the Hund's rule coupling and the extended s-wave orbital order is disfavored by the stripy magnetic correlation pattern in the IBSs.    
\end{abstract}

\pacs{}

\maketitle
Breaking of the tetragonal symmetry in the paramagnetic phase of the iron-based 
superconductors(IBSs) as observed in various measurements has attracted a lot of interests in the community\cite{Fernandes}. 
These phenomena are generally termed as electronic nematicity since the lattice degree of freedom is found to play only a minor role in them.
The origin of the electronic nematicity and its relation with the superconductivity and magnetism of the system are among
the major unsolved issues in the study of the IBSs. 

The first clear evidence for electronic nematicity in the IBSs comes from transport
measurement, in which it is found that the resistivity of the system breaks the tetragonal symmetry at a 
temperature that is significantly higher than the critical temperature for magnetic 
ordering\cite{Chu1,Chu2,Chu3}. It is then found that the uniform susceptibility and magnetic fluctuation spectrum also breaks the 
tetragonal symmetry in the same temperature range\cite{Kasahara,Dai}. The most direct evidence of 
electronic nematic order comes from ARPES measurements\cite{Shimojima,Shen,Zhang,Lu}, in which it was found that 
the degeneracy between the $3d_{yz}$-dominated band around the $\mathrm{M}$ point($\mathrm{k}=[\pi,0]$) and 
the $3d_{xz}$-dominated band around the $\mathrm{M}'$ point($\mathrm{k}=[0,\pi]$) is already lifted in the paramagnetic phase. 

Early theories suggest that the observed electronic nematicity may be related to the stripy magnetic ordering pattern of the system\cite{Fang,Xu,Chubukov1}. More specifically, the tetragonal symmetry breaking involved in the stripy magnetic ordering may survive the magnetic phase transition, resulting in a paramagnetic phase with anisotropic spin correlation. Such a spin nematic scenario has been widely used to understand the phase diagram of the IBSs. However, the spin nematic theory does not explain how other electronic nematic behaviors of the IBSs are induced by the spin nematicity. For example, the band splitting observed in ARPES measurements is so large that it is hard to be taken as a by-product of spin nematicity\cite{Shen,Zhang,Lu}. At the same time, the electronic nematicity is not always accompanied with stripy magnetic ordering. For example, it is found recently that in FeSe the electronic nematicity develops even if the system is far from any magnetic ordering instability\cite{FeSe}. 

Orbital order is another widely discussed origin for the electronic nematicity of the IBSs\cite{Ku,Lv1,Singh,Lv2}. The large band splitting observed in ARPES experiments indicates that the orbital order can indeed be taken as the primary order parameter of the electronic nematic phase. However, it is debated what is the driving force of such an orbital order and what is the relation between the orbital order and the spin nematicity\cite{Fernandes}. More recently, it is found that the band splitting in the IBSs exhibits strong momentum dependence\cite{Zhang,Lu}. In particular, the band splitting around the $\mathrm{M/M'}$ point is found to be much more pronounced than that around the $\Gamma$ point. At the same time, while the band splitting around the $\mathrm{M/M'}$ point exhibits order-parameter-like temperature dependence below the nematic transition point, the band splitting around the $\Gamma$ point is almost temperature independent and persists far above the nematic transition point. It is suggested that such a peculiar momentum dependence can be understood by assuming a d-wave form factor for the orbital order\cite{Su1,Su2} and attributing the band splitting around the $\Gamma$ point to the spin-orbital coupling effect\cite{Zhang}. However, it is not clear why the orbital order in the IBSs chooses such a special form\cite{Jiang,Tao,Christensen,Fanfarillo}.

We think the key to understand the origin of the electronic nematicity in the IBSs is to note the strong entanglement between the orbital and the magnetic degree of freedom in such a multi-orbital system\cite{Nevidomskyy,Su2,Su3,Chubukov2}. In general, the orbital degree of freedom is indispensable in the description of the magnetic property of a multi-orbital system\cite{Kuroki1,Kuroki2,Christensen,Chubukov2}, since electrons in different orbital may experience electron correlation of different strengths and may favor magnetic correlation of different patterns. Thus, the breaking of the spin rotational symmetry in the magnetic ordered phase of a multi-orbital system will generally be accompanied by symmetry breaking in the orbital space. One should better take both the magnetic and the orbital order as components of a composite order parameter. This point is perfectly illustrated in the IBSs, in which the symmetry related $3d_{xz}$ and $3d_{yz}$ orbital prefer stripy magnetic order of wave vector $\mathrm{Q}=(0,\pi)$ and $\mathrm{Q'}=(\pi,0)$ respectively. The orbital order that breaks the tetragonal symmetry relating the $3d_{xz}$ and $3d_{yz}$ orbital will couple linearly to the stripy magnetic order and will be generated spontaneously in the magnetic ordered phase\cite{Nevidomskyy,Su2}. This can actually be understood as a generalized Jahn-Teller effect, in which the degenerate magnetic ordering pattern with wave vector $\mathrm{Q}$ and $\mathrm{Q'}$ play the role of the degenerate electronic states in the case of the conventional Jahn-Teller effect, while the orbital order plays the role of the local distortion.

The above scenario can be naturally extended to the paramagnetic phase, in which it is the difference in the magnetic correlation around the wave vector $\mathrm{Q}$ and $\mathrm{Q'}$ that are coupled linearly to the orbital order. As a result of such a coupling, additional gain in the magnetic correlation energy is expected when a finite orbital order is established in the system. We take this as the driving force of orbital order in the IBSs. We emphasis that the orbital order should not be understood as a by-product of the spin nematicity, but should be understood as an essential component of the nematic order that is driven together with the spin nematicity.

To illustrate this point, we calculate the magnetic correlation energy in the paramagnetic phase of the IBSs in the presence of the orbital order. The magnetic fluctuation and the correlation energy will be treated in the itinerant picture within the multi-orbital random phase approximation(RPA). We will focus on two points. First, we will investigate if orbital order can indeed result in additional gain in the magnetic correlation energy. Second, we will investigate what kind of orbital order is favored by such a mechanism.

In our study, we will adopt the five band model derived form fitting the first principle band structure of the LaFeAsO system. We believe the mechanism discussed in this paper is universal and insensitive to the choice of the band parameters. The band model reads\cite{Kuroki1,Kuroki2}
\begin{eqnarray}
H_{kin}=\sum_{i,j}\sum_{\mu,\nu,\sigma}[t_{i,j}^{\mu,\nu}c_{i,\mu,\sigma}^{\dagger}c_{j,\nu,\sigma}+h.c.]
 +\sum_{i,\mu,\sigma}\varepsilon_{\mu}n_{i,\mu,\sigma}, \nonumber 
\end{eqnarray}
where $\mu,\nu=1,..,5$ is the index of the five $3d$ orbital on the Fe site, with
$|1\rangle=|3d_{3Z^{2}-R^{2}}\rangle$, $|2\rangle=|3d_{XZ}\rangle$, $|3\rangle=|3d_{YZ}\rangle$, $|4\rangle=|3d_{X^{2}-Y^{2}}\rangle$
and $|5\rangle=|3d_{XY}\rangle$. $t_{i,j}^{\mu,\nu}$ is the hopping integral between the $\mu$-th orbital on site $i$ and the $\nu$-th orbital on site $j$, $\varepsilon_{\mu}$ is the on-site energy of the $\mu$-th orbital. The model parameters are chosen in such a ways as to preserve the tetragonal symmetry of the system. A list of the value of the model parameters can be found in [\onlinecite{Kuroki1,Kuroki2}].

The electron interaction has the usual Hubbard-Kanamori form
\begin{eqnarray}
 &H_{int}&=U\sum_{i,\mu}n_{i,\mu,\uparrow}n_{i,\mu,\downarrow}+U'\sum_{i,\mu>\nu}n_{i,\mu}n_{i,\nu}\nonumber\\
 &-J&\sum_{i,\mu \neq \nu}\vec{\mathrm{S}}_{i,\mu}\cdot \vec{\mathrm{S}}_{i,\nu}+J\sum_{i,\mu \neq
 \nu}c^{\dagger}_{i,\mu,\uparrow}c^{\dagger}_{i,\mu,\downarrow}c_{i,\nu,\downarrow}c_{i,\nu,\uparrow}.\nonumber
\end{eqnarray}
Here we have included the intra-orbital and the inter-orbital Coulomb repulsion, the Hund's rule coupling and the pair hopping term.  $n_{i,\mu}$ and $\vec{\mathrm{S}}_{i,\mu}$ are the electron number and spin operator on the $\mu$-th orbital of site $i$. In our study we set $U=1.2\ eV$, $U'=0.9\ eV$ and $J=0.15\ eV$, as has been done in Ref[\onlinecite{Kuroki1,Kuroki2}] .

To describe the effect of the spontaneously generated orbital order, we introduce the following phenomenological term 
\begin{eqnarray}
 H_{\eta}=\eta \hat{O}+ \alpha \eta^{2},\nonumber
\end{eqnarray}
in which $\hat{O}$ denotes the operator of the orbital order. The last term $\alpha\eta^{2}$ is a restoring force term introduced to ensure the stability of the tetragonal phase when the effect of the magnetic correlation energy can be neglected. In a previous work, we have shown that up to the nearest neighboring bonds, the orbital order in the electronic nematic phase of the IBSs can only take three forms\cite{Su1,Su2}. The three forms are, (1) an on-site form that distinguish the energy of the $3d_{xz}$ and the $3d_{yz}$ orbital\cite{Rotation}, which is given by
\begin{eqnarray}
\hat{O}_{on-site}=\sum_{i,\sigma}(c^{\dagger}_{i,2,\sigma}c_{i,2,\sigma}-c^{\dagger}_{i,3,\sigma}c_{i,3,\sigma})\nonumber
\end{eqnarray}
 (2)an extended s-wave form that distinguish the nearest neighboring hopping integral of the $3d_{xz}$ and $3d_{yz}$ orbital, which is given by
\begin{eqnarray}
\hat{O}_{s-wave}=\sum_{i,\delta,\sigma}(c^{\dagger}_{i+\delta,2,\sigma}c_{i,2,\sigma}-c^{\dagger}_{i+\delta,3,\sigma}c_{i,3,\sigma})\nonumber
\end{eqnarray}
and (3) a d-wave form that distinguish the nearest neighboring hopping integral of both orbital in the x- and the y-direction, which is given by
\begin{eqnarray}
 \hat{O}_{d-wave}=\sum_{i,\delta,\sigma}\mathrm{d}_{\delta}(c^{\dagger}_{i+\delta,2,\sigma}c_{i,2,\sigma}
+c^{\dagger}_{i+\delta,3,\sigma}c_{i,3,\sigma}).\nonumber
\end{eqnarray} 
Here $\delta=\pm x,\pm y$ is the vector connecting nearest neighboring Fe sites. $\mathrm{d}_{\delta}$ is the d-wave form  
factor and is given by $\mathrm{d}_{\pm x}=1$, $\mathrm{d}_{\pm y}=-1$. $\eta$ denotes the magnitude of the orbital order.

We now study the effect of the orbital order on the magnetic correlation energy of the system, which will be denoted as $E_{mag}(\eta)$. In the magnetic ordered phase, $E_{mag}(\eta)$ should be linear in $|\eta|$ for small $\eta$, since the tetragonal symmetry is already broken by the stripy magnetic order, which act as a linear bias in the degenerate space of the $3d_{xz}$ and the $3d_{yz}$ orbital. On the other hand, in the paramagnetic phase, $E_{mag}(\eta)$ should be quadratic in $\eta$ for small $\eta$ as a result of the tetragonal symmetry. $E_{mag}(\eta)$ can be calculated from the dynamical spin susceptibility of the system. For a multi-orbital system, the dynamical spin susceptibility is defined as
\begin{eqnarray}
\boldsymbol{\chi}_{\mu\nu,\mu'\nu'}(\mathrm{q},\tau)=-<\mathrm{T}_{\tau} \mathrm{S}_{\mu\nu}(\mathrm{q},\tau) \ \mathrm{S}^{\dagger}_{\mu'\nu'}(\mathrm{q},0)>,\nonumber
\end{eqnarray}
in which  $\mathrm{S}_{\mu\nu}(\mathrm{q})=\frac{1}{2}\sum_{\mathrm{k}}(c^{\dagger}_{\mathrm{k+q},\mu,\uparrow}c_{\mathrm{k},\nu,\uparrow}-c^{\dagger}_{\mathrm{k+q},\mu,\downarrow}c_{\mathrm{k},\nu,\downarrow})$ is the Fourier component of the spin density operator at momentum $\mathrm{q}$. Here we use bold symbols to denote matrix objects and $\boldsymbol{\chi}$ is a $25\times25$ matrix for a five-band system. The orbital character of the magnetic fluctuation can be found from the spectral representation of $\boldsymbol{\chi}$, which is given by
\begin{eqnarray}
\boldsymbol{\chi}_{\mu\nu, \mu'\nu'}(\mathrm{q},i\omega_{n})=\int_{-\infty}^{\infty}\frac{d\omega'}{2\pi}\frac{\mathrm{\mathbf{ R}}_{\mu\nu,\mu'\nu'}(\mathrm{q},\omega')}{i\omega_{n}-\omega'}.\nonumber
\end{eqnarray} 
Here the matrix elements of the spectral density matrix $\mathrm{\mathbf{R}}(\mathrm{q},\omega)$ is given by
\begin{eqnarray}
\mathrm{\mathbf{R}}_{\mu\nu, \mu'\nu'}(\mathrm{q},\omega)= \sum_{m,n}<n|\mathrm{S}_{\mu\nu}(\mathrm{q})|m><m|\mathrm{S}^{\dagger}_{\mu'\nu'}(\mathrm{q})|n>\nonumber\\
\times 2\pi e^{\beta\Omega}(e^{-\beta E_{n}}-e^{-\beta E_{m}})\delta(\omega+E_{n}-E_{m}),\nonumber
\end{eqnarray}
in which $E_{n}$ and $E_{m}$ are the eigen-energies of the system. From this expression, we see $\mathrm{\mathbf{R}}(\mathrm{q},\omega)$ is a Hermitian matrix. The eigenvalues and eigenvectors of $\mathrm{\mathbf{R}}(\mathrm{q},\omega)$ can thus be interpreted as the spectral weight and the orbital character of the magnetic fluctuation. The spectral density matrix $\mathrm{\mathbf{R}}(\mathrm{q},\omega)$ can be found from the retarded dynamical spin susceptibility as follows
\begin{equation}
\mathrm{\mathbf{R}}(\mathrm{q},\omega)=i[\boldsymbol{\chi}(\mathrm{q},\omega+i0^{+})-\boldsymbol{\chi}^{\dagger}(\mathrm{q},\omega+i0^{+})].\nonumber
\end{equation}
As usual, the spin structure factor of the system is related to the spectral density through the following sum rule
\begin{equation}
<\mathrm{S}_{\mu\nu}(\mathrm{q}) \ \mathrm{S}_{\nu'\mu'}(-\mathrm{q})>=\int_{0}^{\infty} d\omega \ \mathrm{\mathbf{ R}}_{\mu\nu,\mu'\nu'}(\mathrm{q},\omega).\nonumber
\end{equation}

In the RPA scheme, the dynamical spin susceptibility is given by
\begin{eqnarray}
\boldsymbol{\chi}(\mathrm{q},i\omega_{n})=\frac{\boldsymbol{\chi}_{0}(\mathrm{q},i\omega_{n})}{\mathrm{\mathbf{I}}-\mathrm{\mathbf{V}}\boldsymbol{\chi}_{0}(\mathrm{q},i\omega_{n})}.\nonumber
\end{eqnarray}
This should be understood as a matrix equation in the $25\times25$ orbital space.  $\boldsymbol{\chi}_{0}$ is the bare spin susceptibility of the five band model, whose matrix element is given by
\begin{eqnarray}
[\boldsymbol{\chi}_{0}(\mathrm{q},i\omega_{n})]_{\mu\nu,\ \mu'\nu'}=\sum_{\mathrm{k},m,n}\frac{u^{*}_{\mathrm{k+q},\mu,m}u_{\mathrm{k},\nu,n}u^{*}_{\mathrm{k},\mu',n}u_{\mathrm{k+q},\nu',m}}{i\omega_{n}-(\xi_{\mathrm{k+q},m}-\xi_{\mathrm{k},n})}.\nonumber
\end{eqnarray}
Here, $\xi_{n}(\mathrm{k})$ denotes the $n$-th eigenvalue of the band Hamiltonian at momentum $\mathrm{k}$, $u_{\mathrm{k},\mu,n}$ denotes the $n$-th eigenvector of the band Hamiltonian at momentum $\mathrm{k}$. $\mathrm{\mathbf{V}}$ is the interaction kernel in the magnetic channel. For our model its matrix element is given by\cite{Kuroki2}
\begin{eqnarray}
V_{\mu\nu, \ \mu'\nu'}=\left\{\begin{array}{lr} U&\mu=\nu,\mu'=\nu', \mu=\mu'\\ U'& \mu=\mu',\nu=\nu',\mu\neq\nu \\J & \mu=\nu,\mu'=\nu',\mu\neq\mu'\\J&\mu=\nu',\nu=\mu',\mu\neq\nu\end{array}\right.\nonumber
\end{eqnarray}

The correlation energy related to the magnetic fluctuation can be found from the interaction integral method and is given by
\begin{eqnarray}
E_{mag}(\eta)&=&\ \ \  \int_{0}^{1}d\lambda <H_{int}>_{\lambda}\nonumber\\
&=&-\frac{3}{2}\int_{0}^{1}d\lambda\int_{0}^{\infty}d\omega \sum_{\mathrm{q}} \mathrm{Tr}[\mathrm{\mathbf{V}}\mathrm{\mathbf{R}}^{\lambda}(\mathrm{q},\omega)]. \nonumber
\end{eqnarray} 
Here  $\mathrm{\mathbf{R}}^{\lambda}(\mathrm{q},\omega)$ is the spectral density matrix calculated at the interaction strength $\lambda H_{int}$. Note that here we have only retained the interaction term in the magnetic channel, which is approximated by 
\begin{equation}
H_{mag}=-\frac{1}{2}\sum_{\mathrm{q},\mu\nu,\mu'\nu'}V_{\mu\nu,\mu'\nu'}\vec{\mathrm{S}}_{\mu\nu}(\mathrm{q})\cdot\vec{\mathrm{S}}_{\nu'\mu'}(-\mathrm{q}).\nonumber
\end{equation}
The logic behind this approximation is that the remaining part of $H_{int}$ is less sensitive to the presence of the orbital order. 

\begin{figure}
\includegraphics[width=8cm]{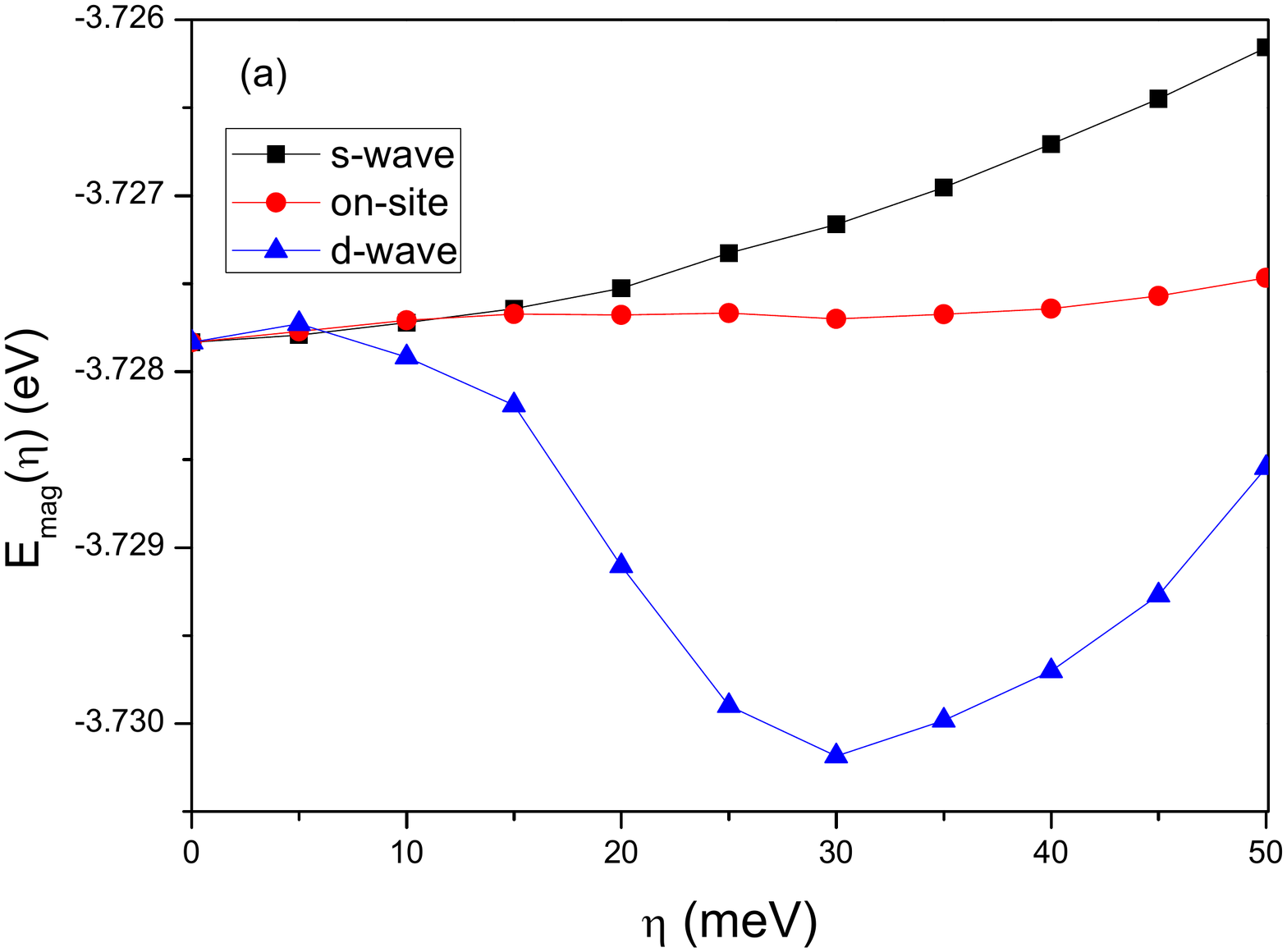}
\includegraphics[width=8cm]{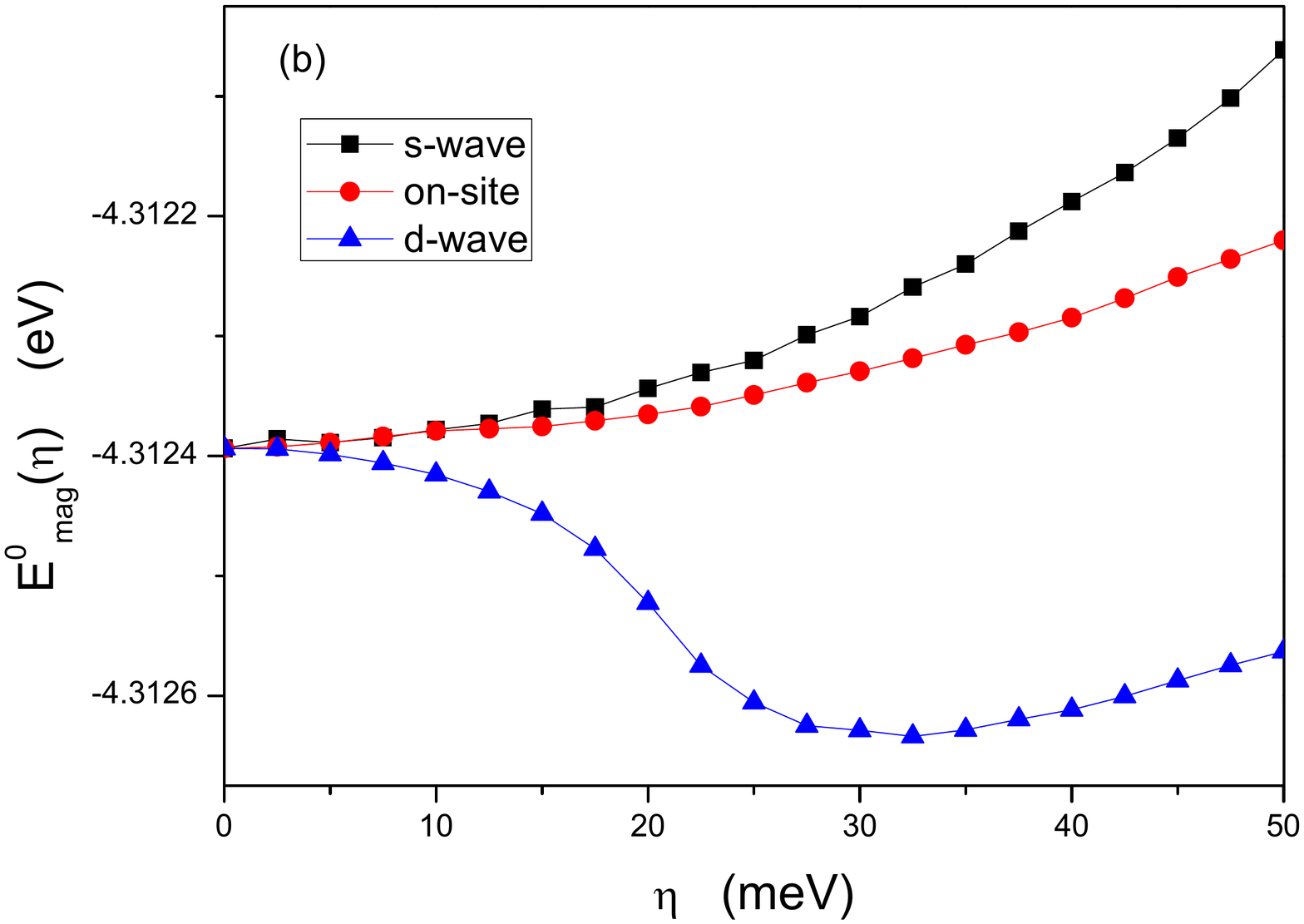}
\caption{\label{fig1}
(Color on-line) (a)The magnetic correlation energy in the presence of the orbital order of the on-site, extended s-wave and d-wave form. Here $\lambda=0.8$. (b)The mean value of $H_{mag}$ in the non-interacting ground state for the three forms of orbital order. }
\end{figure}

In our study, we focus on the ground state property and treat the strength of the interaction as a tuning parameter. For electron density $n=6$, we find that the magnetic instability occurs when $\lambda>\lambda_{c}\simeq 0.85$ at $\eta=0$. When $\eta\neq0$, the value of $\lambda_{c}$ for stripy magnetic order with wave vector $\mathrm{Q}$ and $\mathrm{Q'}$ will be different. More specifically, the $\lambda_{c}$ for the favored wave vector will be reduced and that for the disfavored wave vector will be enhanced\cite{Nevidomskyy,Su2}.

Our calculation is done on a $96\times 96$ lattice. The result for $E_{mag}(\eta)$ is plotted in Figure 1a. Here the electron density is fixed at $n=6$ and the interaction strength is fixed at $\lambda=0.8$, slightly below the critical value for magnetic instability. One find that among the three forms of orbital order, only the d-wave orbital order can improve the magnetic correlation energy of the system\cite{Nematic}. This result can be understood from a mean field analysis of the magnetic interaction. In Figure 1b, we plotted the mean value of $H_{mag}$ in the non-interacting ground state with $n=6$, which is given by 
\begin{eqnarray}
E^{0}_{mag}(\eta)=\frac{3(U-U')}{16}\sum_{\mu}n^{2}_{\mu}+\frac{3J}{8}\sum_{\mu\neq\nu}\chi^{2}_{\mu,\nu}-\frac{9(U+U')}{4}.\nonumber
\end{eqnarray}
Here $n_{\mu}=\sum_{\sigma}\expval{c^{\dagger}_{i,\mu,\sigma}c_{i,\mu,\sigma}}$, $\chi_{\mu,\nu}=\sum_{\sigma}\expval{c^{\dagger}_{i,\mu,\sigma}c_{i,\nu,\sigma}}$. It is found that the qualitative feature of the RPA calculation is well captured by the mean field analysis, albeit with a much reduced scale in the variation of the magnetic correlation energy as we vary $\eta$. 

To look more closely into the reason why the d-wave orbital order is favored over the other two forms of orbital order, we compare the part of $E^{0}_{mag}$ that is proportional to $U$, $U'$ and $J$, which will be denoted as $E^{0}_{U}$, $E^{0}_{U'}$ and $E^{0}_{J}$ in the following. Form Figure 2a, we see that both the on-site and d-wave orbital order can significantly improve $E^{0}_{U}+E^{0}_{U'}$, while the s-wave orbital order is slightly harmful to $E^{0}_{U}+E^{0}_{U'}$\cite{Correlation}. The origin for such a contrasting behavior can be found in the momentum dependence of the band splitting induced by the three forms of orbital order. Unlike the on-site and the d-wave orbital order, the band splitting induced by the extended s-wave orbital order is strongly suppressed around the $\mathrm{M/M'}$ point, which is the most important momentum region for stripy magnetic ordering with wave vector $\mathrm{Q/Q'}$. The sensitivity of the magnetic correlation energy to the band splitting at the $\mathrm{M/M'}$ point is thus a manifestation of the entanglement between the orbital order and the spin nematicity. The situation for $E^{0}_{J}$ is totally different. From the plot in Figure 2b, one find that the on-site orbital order will significantly increase $E^{0}_{J}$, while the increase of $E^{0}_{J}$ related to the d-wave and the extended s-wave orbital order is much less pronounced. This is reasonable since the on-site orbital order acts effectively as a crystal field splitting and competes directly with the Hund's rule coupling. These considerations leave the d-wave orbital order as the only possibility for the IBSs.

The size of the orbital order is determined by both the magnetic correlation energy $E_{mag}(\eta)$ and the restoring force $\alpha\eta^{2}$, which is a prior unknown to us. 
Nevertheless, we can use the minimum point of $E_{mag}(\eta)$ as an estimate of the size of the orbital order. For the parameters that we have used, the band splitting at the M point is about $8\eta_{min}\simeq240 \ \mathrm{meV}$, which is about three times larger than that observed in experiment\cite{Shen,Zhang,Lu}. However, the actual band width of the IBSs is also about three times smaller than that predicted by first principle calculation\cite{Shen,Zhang,Lu}. 

Finally, since the gain in the magnetic correlation energy will diverge as the system approaches the magnetic instability\cite{Divergence}, the orbital ordering must happen somewhere before the stripy magnetic ordering transition. We note the RPA scheme usually underestimates the strength of the magnetic fluctuation in the paramagnetic phase, for which a more accurate treatment of the electron correlation effect is required. We thus believe that the electronic nematic phase should be more robust than that predicted by the RPA theory and may not necessarily be accompanied with stripy magnetic ordering transition.

In conclusion, we find that the magnetic correlation energy favors a d-wave orbital order over the on-site and the extended s-wave orbital order in the electronic nematic phase of the IBSs as a result of the stripy magnetic correlation pattern and the Hund's rule coupling. We find that the orbital order in the IBSs should not be understood as a by-product of spin nematicity, but is an essential component of a composite order parameter involving both the magnetic and the orbital degree of freedom.

\begin{figure}
\includegraphics[width=8cm]{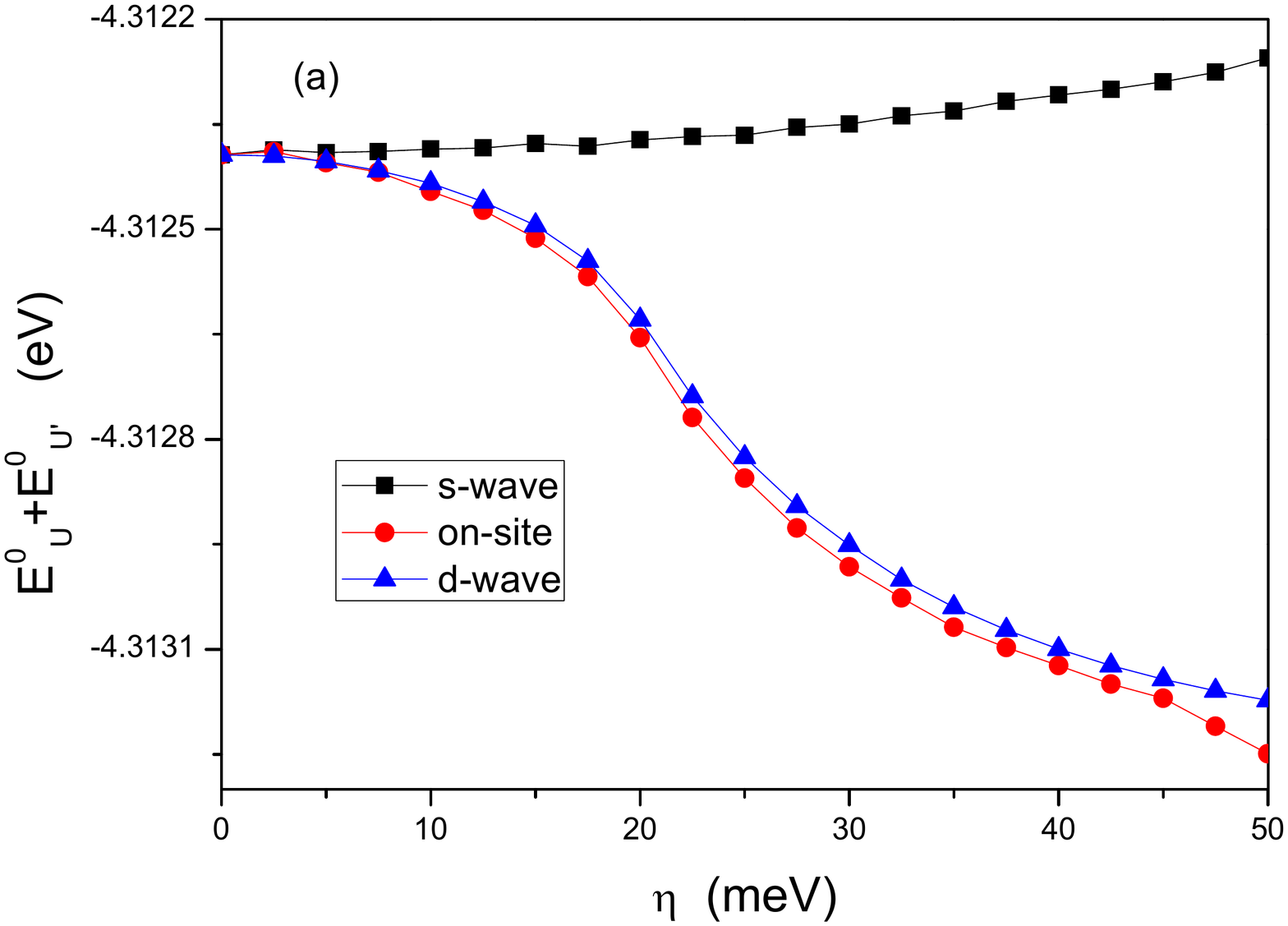}
\includegraphics[width=8cm]{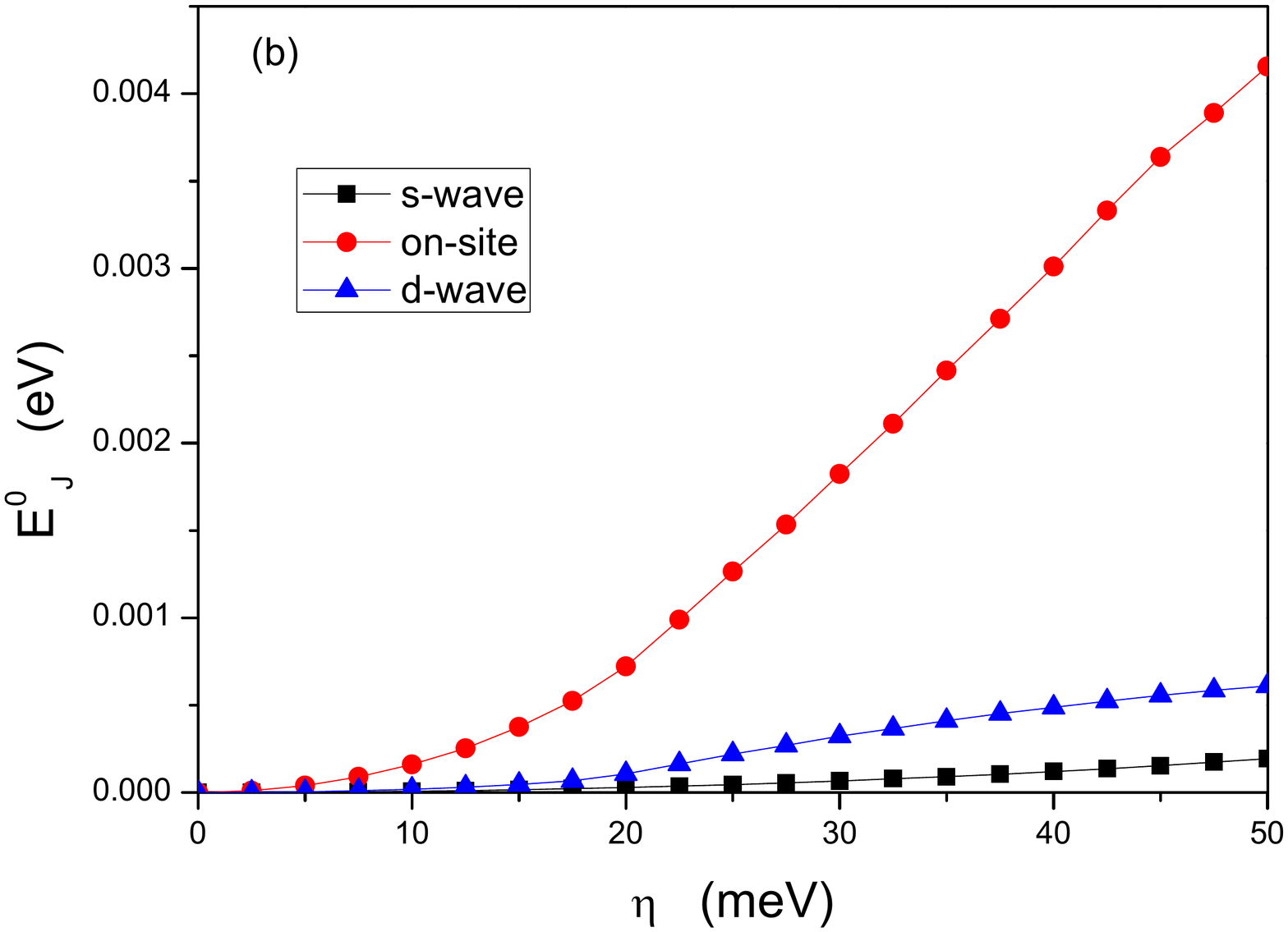}
\caption{\label{fig3}
(Color on-line) The mean value of $H_{mag}$ in the non-interacting ground state. (a) $E^{0}_{U}+E^{0}_{U'}$, (b)$E^{0}_{J}$. Here we have rescaled $\eta$ by a factor of 4 in the case of the on-site orbital order so that the maximal band splitting in the Brillouin zone are the same for the three forms of orbital order.}
\end{figure}


\begin{thebibliography}{99}
\bibitem{Rotation}We note that the principle axis $X$ and $Y$ defined in [\onlinecite{Kuroki1,Kuroki2}] are along the Fe-As direction. For convenience's sake, we will use $|2\rangle$ and $|3\rangle$ to denote $|3d_{xz}\rangle$ and $|3d_{yz}\rangle$. The principle axis $x$ and $y$ are along the Fe-Fe direction.
\bibitem{Nematic}We find that the tetragonal phase with $\eta=0$ is always locally stable with respect to the d-wave orbital order. It is thus incorrect to use the nematic susceptibility to judge the stability of the tetragonal phase.
\bibitem{Correlation}Since $E^{0}_{U}$ and $E^{0}_{U'}$ anti-correlate with each other as we vary $\eta$, we present the result of $E^{0}_{U}+E^{0}_{U'}$, rather than the results of $E^{0}_{U}$ and $E^{0}_{U'}$ separately.
\bibitem{Divergence}Such a divergence will be eventually cut off by corrections beyond the RPA when the system is close enough to the magnetic transition point, acrossing which $E_{mag}(\eta)$ becomes linear in $|\eta|$ for small $\eta$.
\end{thebibliography}
\end{document}